\documentstyle[aasms4]{article}
\makeatletter
  
  \@addtoreset{equation}{section}
\makeatother

\begin{document}

\title{RELATIVISTIC CORRECTIONS TO THE SUNYAEV-ZEL'DOVICH EFFECT FOR CLUSTERS OF GALAXIES. V. MULTIPLE SCATTERING}

\author{NAOKI ITOH AND YOUHEI KAWANA}

\affil{Department of Physics, Sophia University, 7-1 Kioi-cho, Chiyoda-ku, Tokyo, \\
102-8554, Japan; n\_itoh, y-kawana@hoffman.cc.sophia.ac.jp}

\author{SATOSHI NOZAWA}

\affil{Josai Junior College for Women, 1-1 Keyakidai, Sakado-shi, Saitama, \\
350-0295, Japan; snozawa@galaxy.josai.ac.jp}

\centerline{AND}

\author{YASUHARU KOHYAMA}

\affil{Fuji Research Institute Corporation, 2-3 Kanda-Nishiki-cho, Chiyoda-ku, Tokyo, \\
101-8443, Japan; kohyama@star.fuji-ric.co.jp}

\begin{abstract}

  We extend the formalism for the calculation of the relativistic corrections to the 
Sunyaev-Zel'dovich effect for clusters of galaxies and include the multiple scattering effects.  We present a systematic method for the inclusion of the multiple scattering effects.  The multiple scattering contribution is found to be very small compared with the single scattering contribution.  For high-temperature galaxy clusters of $k_{B} T_{e} \approx 15$keV, the ratio of the both contributions is $-0.3\%$ in the Wien region.  In the Rayleigh--Jeans region the ratio is  $-0.03\%$.  Therefore the multiple scattering contribution is safely neglected for the observed galaxy clusters.

\end{abstract}


\section{INTRODUCTION}

  Compton scattering of the cosmic microwave background (CMB) radiation by hot intracluster gas --- the Sunyaev-Zel'dovich effect (Zel'dovich \& Sunyaev 1969; Sunyaev \& Zel'dovich 1972, 1980a, 1980b, 1981) --- provides a useful method to measure the Hubble constant $H_{0}$ (Gunn 1978; Silk \& White 1978; Birkinshaw 1979; Cavaliere, Danese, \& De Zotti 1979; Birkinshaw, Hughes, \& Arnaud 1991; Birkinshaw \& Hughes 1994; Myers et al. 1995; Herbig et al. 1995; Jones 1995; Markevitch et al. 1996; Holzapfel et al. 1997; Furuzawa et al. 1998; Komatsu et al. 1999).  The original Sunyaev-Zel'dovich formula has been derived from a kinetic equation for the photon distribution function taking into account the Compton scattering by electrons: the Kompaneets equation (Kompaneets 1957; Weymann 1965).  The original Kompaneets equation has been derived with a nonrelativistic approximation for the electron.  However, recent X-ray observations have revealed the existence of many high-temperature galaxy clusters (David et al. 1993; Arnaud et al. 1994; Markevitch et al. 1994; Markevitch et al. 1996; Holzapfel et al. 1997; Mushotzky \& Scharf 1997; Markevitch 1998).  In particular, Tucker et al. (1998) reported the discovery of a galaxy cluster with the electron temperature $k_{B} T_{e} = 17.4 \pm 2.5$ keV.  Rephaeli and his collaborator (Rephaeli 1995; Rephaeli \& Yankovitch 1997) have emphasized the need to take into account the relativistic corrections to the Sunyaev-Zel'dovich effect for clusters of galaxies.

  In recent years remarkable progress has been achieved in the theoretical studies of the relativistic corrections to the Sunyaev-Zel'dovich effects for clusters of galaxies.  Stebbins (1997) generalized the Kompaneets equation.  Itoh, Kohyama, \& Nozawa (1998) have adopted a relativistically covariant formalism to describe the Compton scattering process (Berestetskii, Lifshitz, \& Pitaevskii 1982; Buchler \& Yueh 1976), thereby obtaining higher-order relativistic corrections to the thermal Sunyaev-Zel'dovich effect in the form of the Fokker-Planck expansion.  In their derivation, the scheme to conserve the photon number at every stage of the expansion which has been proposed by Challinor \& Lasenby (1998) played an essential role.  The results of Challinor \& Lasenby (1998) are in agreement with those of Itoh, Kohyama, \& Nozawa (1998).  The latter results include higher-order expansions.  Itoh, Kohyama, \& Nozawa (1998) have also calculated the collision integral of the Boltzmann equation numerically and have compared the results with those obtained by the Fokker-Planck expansion method.  They have confirmed that the Fokker-Planck expansion method gives an excellent result for $k_{B}T_{e} \leq 15$keV, where $T_{e}$ is the electron temperature.  For $k_{B}T_{e} \geq 15$keV, however, the Fokker-Planck expansion results show nonnegligible deviations from the results obtained by the numerical integration of the collision term of the Boltzmann equation.

  Nozawa, Itoh, \& Kohyama (1998b) have extended their method to the case where the galaxy cluster is moving with a peculiar velocity with respect to CMB.  They have thereby obtained the relativistic corrections to the kinematical Sunyaev-Zel'dovich effect.  Challinor \& Lasenby (1999) have confirmed the correctness of the result obtained by Nozawa, Itoh, \& Kohyama (1998b).  Sazonov \& Sunyaev (1998a, b) have calculated the kinematical Sunyaev-Zel'dovich effect by a different method.  Their results are in agreement with those of Nozawa, Itoh, \& Kohyama (1998b).  The latter authors have given the results of the higher-order expansions.

  Itoh, Nozawa, \& Kohyama (2000) have also applied their method to the calculation of the relativistic corrections to the polarization Sunyaev-Zel'dovich effect (Sunyaev \& Zel'dovich 1980b, 1981).  They have thereby confirmed the result of Challinor, Ford, \& Lasenby (1999) which has been obtained with a completely different method.  Recent works on the polarization Sunyaev-Zel'dovich effect include Audit \& Simons (1999), Hansen \& Lilje (1999), and Sazonov \& Sunyaev (1999).

  As stated above, Itoh, Kohyama, \& Nozawa (1998) have carried out the numerical integration of the collision term of the Boltzmann equation.  This method produces the exact results without the power series expansion approximation.  Sazonov \& Sunyaev (1998a, b) have reported the results of the Monte Carlo calculations on the relativistic corrections to the Sunyaev-Zel'dovich effect.  In Sazonov \& Sunyaev (1998b), a numerical table which summarizes the results of the Monte Carlo calculations has been presented.  This table is of great value when one wishes to calculate the relativistic corrections to the Sunyaev-Zel'dovich effect for galaxy clusters of extremely high temperatures.  Accurate analytic fitting formulae would be still more convenient to use for the observers who wish to analyze the galaxy clusters with extremely high temperatures.  For this purpose, Nozawa et al. (2000) have presented an accurate analytical fitting formula of 0.1\% accuracy for the numerical results for the relativistic corrections to the thermal Sunyaev-Zel'dovich effect for clusters of galaxies.  For the analyses of the galaxy clusters with extremely high temperatures, the results of the calculation of the relativistic thermal bremsstrahlung Gaunt factor (Nozawa, Itoh, \& Kohyama 1998a) and their accurate analytic fitting formulae (Itoh et al. 2000) will be useful.

  In this series of papers devoted to the study of the relativisitc corrections to the Sunyaev-Zel'dovich effect for clusters of galaxies, we have so far restricted ourselves to the case of single Compton scattering.  This is justified because the optical depth for the Compton scattering of the CMB photon inside the galaxy clusters is generally about $10^{-2}$ or smaller (Birkinshaw 1999).  Nevertheless, it would be desirable to evaluate the effects of the multiple Compton scattering of the CMB photon inside the galaxy clusters accurately, as we have already developed the method to calculate the relativistic corrections to the Sunyaev-Zel'dovich effect for the galaxy clusters with high accuracy.  The multiple scattering effects have been already considered by many authors (see Birkinshaw 1999 for references).  In this paper we wish to evaluate the multiple scattering effects in the same theoretical framework of this series of papers.

  The present paper is organized as follows.  In $\S$ 2 we give the method of the calculation and the results.  In $\S$ 3 we give discussion of the results and concluding remarks.

\section{MULTIPLE SCATTERING CONTRIBUTION}

  In the present paper, we would like to derive the analytic expression for the multiple scattering contribution in the Sunyaev-Zeldovich effect for clusters of galaxies.  As a reference system, we choose the system that is fixed to the center of mass of the galaxy cluster.  The galaxy cluster is assumed to be fixed to the cosmic microwave background (CMB).  Following Itoh, Kohyama \& Nozawa (1998), we start with the Fokker-Plank expansion for the time evolution equation of the CMB photon distribution function $n(\omega)$:
\begin{eqnarray}
\frac{ \partial n(\omega)}{ \partial t} & = & 
 2 \left[ \frac{ \partial n}{ \partial x} + n(1+n) \right] \, I_{1} 
  \nonumber  \\
& + & 2 \left[ \frac{ \partial^{2} n}{ \partial x^{2}} 
+ 2(1+n) \frac{ \partial n}{ \partial x} + n(1+n)  \right] \, I_{2}
  \nonumber  \\
& + & 2 \left[\frac{ \partial^{3} n}{ \partial x^{3}} 
+ 3(1+n) \frac{ \partial^{2} n}{ \partial x^{2}} 
+ 3(1+n) \frac{ \partial n}{ \partial x} + n(1+n)  \right] \, I_{3}
  \nonumber \\
& + & 2 \left[\frac{ \partial^{4} n}{ \partial x^{4}} 
+ 4(1+n) \frac{ \partial^{3} n}{ \partial x^{3}}
+ 6(1+n) \frac{ \partial^{2} n}{ \partial x^{2}}
+ 4(1+n) \frac{ \partial n}{ \partial x} + n(1+n)  \right] \, I_{4}
  \nonumber \\
& + & \cdot \cdot \cdot  \, \, \, ,
\end{eqnarray}
where 
\begin{eqnarray}
x  & \equiv & \frac{\omega}{k_{B} T_{e}}  \, ,  \\
\Delta x  & \equiv & \frac{ \omega^{\prime} - \omega}{k_{B} T_{e}}  \, ,  \\
I_{k} & \equiv &  \frac{1}{k !} \int \frac{d^{3}p}{(2\pi)^{3}} d^{3}p^{\prime} d^{3}k^{\prime} \, W \, f(E) \, (\Delta x)^{k} \, .
\end{eqnarray}
In equation (2.4), $W$ is the transition probability of the Compton scattering, $f(E)$ is the relativistic Maxwellian distribution function for electrons with temperature $T_{e}$.
We have integrated equation (2.4) analytically with power series expansions of the integrand.  The expansion parameter is 
\begin{equation}
\theta_{e} \equiv \frac{k_{B} T_{e}}{mc^{2}}  \, .
\end{equation}
The explicit forms for $I_{k}$ are given in Itoh, Kohyama \& Nozawa (1998).

  We first assume the initial photon distribution of the CMB radiation to be Planckian with temperature $T_{0}$:
\begin{eqnarray}
n (X) & = & n_{0} (X) \, \equiv \, \frac{1}{e^{X} - 1} \, ,
\end{eqnarray}
where
\begin{eqnarray}
X & \equiv & \frac{\omega}{k_{B} T_{0}}  \, .
\end{eqnarray}
Assuming $T_{0}/T_{e} \ll 1$, one obtains the following expression for the fractional distortion of the photon spectrum derived by Itoh, Kohyama \& Nozawa (1998):
\begin{eqnarray}
\frac{\Delta n(X)}{n_{0}(X)} & = & \frac{y \, \theta_{e} X e^{X}}{e^{X}-1} \, \left[  \,
Y_{0} \, + \, \theta_{e} Y_{1} \, + \, \theta_{e}^{2} Y_{2} \, + \,  \theta_{e}^{3} Y_{3} \, + \,  \theta_{e}^{4} Y_{4} \, \right]  \, ,
\end{eqnarray}
\begin{eqnarray}
y & \equiv & \sigma_{T} \int_{0}^{\ell} d \ell_{1} N_{e}(\ell_{1})  \, ,
\end{eqnarray}
where $\sigma_{T}$ is the Thomson scattering cross section, $N_{e}$ is the electron number density, and the integral is over the photon path length in the cluster.  The explicit forms for $Y_{0}, Y_{1}, Y_{2}, Y_{3}$ and $Y_{4}$ are given in Itoh, Kohyama \& Nozawa (1998).  Equation (2.8) is the single scattering contribution, i.e. the first-order term in $y$.  If the cluster of galaxies is optically thin, i.e. $y \ll 1$, the single scattering approximation is a good approximation.  In fact, the approximation is valid for most of the clusters.  However, it is extremely important to calculate the next-order contribution in order to obtain more accurate theoretical prediction for the future observation of the Sunyaev-Zeldovich effect for clusters of galaxies.

  We now calculate the multiple scattering contribution.  Since $y \ll 1$ is realized for most of clusters of galaxies, the second-order contribution is considered to be sufficient.  We now assume that the initial photon distribution has a first-order perturbation.  Namely,
\begin{eqnarray}
n (X) & = & n_{1} (X) \, \equiv \, n_{0}(X) + \Delta n(X) \, , \nonumber  \\
      &   & \, \, \hspace{1.0cm} \, \, = \, n_{0}(X) \left\{ 1 \, + \, \frac{\Delta n(X)}{n_{0}(X)} \right\}  \, ,
\end{eqnarray}
where the second term in equation (2.10) is given by equation (2.8).  Inserting equation (2.10) into RHS of equation (2.1), and performing the standard calculation, we obtain the following expression for the fractional distortion of the photon distribution function including the second-order contribution:

\begin{eqnarray}
\frac{\Delta n(X)}{n_{0}(X)} & = & \frac{y \, \theta_{e} X e^{X}}{e^{X}-1} \, \left[  \,
Y_{0} \, + \, \theta_{e} Y_{1} \, + \, \theta_{e}^{2} Y_{2} \, + \,  \theta_{e}^{3} Y_{3} \, + \,  \theta_{e}^{4} Y_{4} \, \right]  \, ,  \nonumber  \\
& + &  \frac{1}{2} \, \frac{y^2 \, \theta_{e}^2 X e^{X}}{e^{X}-1} \, \left[  \,
 Z_{0} \, + \, \theta_{e} Z_{1} \, + \, \theta_{e}^{2} Z_{2} \, \right]  \, ,   \\
  \nonumber  \\
Z_{0} & = & - 16 + 34 \tilde{X} - 12 \tilde{X}^2 + \tilde{X}^3 
             \, + \, \tilde{S}^2 \left( - 6 + 2 \tilde{X} \right)  \, ,  \\ 
Z_{1} & = &  - 80 + 590 \tilde{X} - \frac{3492}{5} \tilde{X}^2 + \frac{1271}{5} \tilde{X}^3 
                 - \frac{168}{5} \tilde{X}^4 + \frac{7}{5} \tilde{X}^5   \,  \nonumber  \\ 
      & + &  \tilde{S}^2 \left( -\frac{1746}{5} + \frac{2542}{5} \tilde{X} 
                 - \frac{924}{5} \tilde{X}^2 + \frac{91}{5} \tilde{X}^3 \right) \, \nonumber \\
      & + &  \tilde{S}^4 \left( -\frac{168}{5} + \frac{119}{10} \tilde{X} \right)  \,  \\ 
Z_{2} & = & - 160 + 4792 \tilde{X} - \frac{357144}{25} \tilde{X}^2 
               + \frac{312912}{25} \tilde{X}^3 - \frac{110196}{25} \tilde{X}^4 \, \nonumber \\
      &   &    + \frac{34873}{50} \tilde{X}^5 - \frac{734}{15} \tilde{X}^6 
               + \frac{367}{300} \tilde{X}^7    \,  \nonumber  \\
      & + &   \tilde{S}^2 \left( -\frac{178572}{25} + \frac{625824}{25} \tilde{X} 
              - \frac{606078}{25} \tilde{X}^2 + \frac{453349}{50} \tilde{X}^3 \right.   \,                     \nonumber \\
      &   &   \left. \, \hspace{0.6cm} \, - \frac{20919}{15} \tilde{X}^4 + \frac{367}{5}                \tilde{X}^5 \right)  \,      \nonumber \\       
      & + &   \tilde{S}^4 \left( -\frac{110196}{25} + \frac{592841}{100} \tilde{X} 
                 - 2202 \tilde{X}^2 + \frac{5872}{25} \tilde{X}^3 \right)  \, \nonumber  \\
      & + &   \tilde{S}^6 \left( -\frac{6239}{30} + \frac{11377}{150} \tilde{X} \right) \, , 
\end{eqnarray}
where
\begin{eqnarray}
\tilde{X} & \equiv &  X \, {\rm coth} \left( \frac{X}{2} \right)  \, , \\
\tilde{S} & \equiv & \frac{X}{ \displaystyle{ {\rm sinh} \left( \frac{X}{2} \right)} }   \, .
\end{eqnarray}
In equation (2.11), the first term corresponds to the first-odrder contribution and the second term corresponds to the second-order contribution of the multiple scattering.  In deriving equation (2.11), we have used the following identity relation for $y$:
\begin{eqnarray}
\sigma_{T} \, \int_{0}^{\ell} d \ell_{1} N_{e}(\ell_{1})  \, \sigma_{T} \, \int_{0}^{\ell_{1}} d \ell_{2} N_{e}(\ell_{2}) & = & \frac{1}{2} \left( \sigma_{T} \, \int_{0}^{\ell} d \ell_{1} N_{e}(\ell_{1}) \right)^{2}  \,   \nonumber  \\
& = & \frac{1}{2} \, y^{2}  \, . 
\end{eqnarray}
We have also neglected terms higher than $O(\theta_{e}^4)$ in the $y^{2}$ contributions in equation (2.11).  It is important to note that equation (2.11) satisfies the photon number conservation.  With equation (2.11), we define the distortion of the spectral intensity as follows:
\begin{eqnarray}
\Delta I & = & \frac{X^3}{e^{X} - 1} \, \frac{\Delta n(X)}{n_{0}(X)}  \, = \, \Delta I_{1} \, + \, \Delta I_{2} \, .
\end{eqnarray}
The first term $\Delta I_{1}$ contains a factor $y$, whereas the second term $\Delta I_{2}$ contains a factor $y^{2}$.  In Figure 1 we show $\Delta I_{2}/y^{2}$ as a function of $X$ for the case $k_{B} T_{e}$ = 10keV.  It is clear that the magnitude of $\Delta I_{2}/y^{2}$ has a maximum value at $X \approx 5$ for $k_{B}T_{e}$=10keV.

  In order to estimate the relative importance of the multiple scattering contribution, we now define the following ratio:
\begin{eqnarray}
\Gamma & \equiv & \frac{\Delta I_{2}/y^{2}}{\Delta I_{1}/y}   \, .  
\end{eqnarray}
In Figure 2 we show $\Gamma$ for $X=5$ as a function of the electron temperature $T_{e}$.  It is clear that $\Gamma$ increases with a negative sign as the temperature of the cluster of galaxies increases, i.e., $\Gamma \approx -0.3$ at $k_{B} T_{e}=15$keV.  However, the multiple scattering contribution is small because of a further factor $y$.  Namely, for the cluster of galaxies of $k_{B} T_{e}=15$keV, we have
\begin{eqnarray}
\frac{\Delta I_{2}}{\Delta I_{1}} & = & y \, \Gamma \, \approx -0.3 \, y  \, \approx  -0.3 \% \,  ,
\end{eqnarray}
where we used a typical value $y \approx 0.01$ of the galaxy clusters.  Therefore the maximum effect of the multiple scattering contribution is $-0.3\%$ of the single scattering contribution for the observed high-temperature galaxy clusters.

  In the Rayleigh--Jeans limit where $X \rightarrow 0$, equation (2.11) is further simplified:
\begin{eqnarray}
\frac{\Delta n(X)}{n_{0}(X)} & = & - 2 \, y \, \theta_{e} \, \left( 1 - \frac{17}{10} \theta_{e} + \frac{123}{40} \theta_{e}^{2} - \frac{1989}{280} \theta_{e}^{3}
 + \frac{14403}{640} \theta_{e}^{4} \, \right)  \nonumber  \\
&  &  + 2 \, y^2 \, \theta_{e}^2 \, \left( 1 - \frac{17}{5} \theta_{e} + \frac{226}{25} \theta_{e}^{2} \, \right)  \, .
\end{eqnarray}
With equation (2.21) we have the multiple scattering contribution for $k_{B}T_{e}=$15keV and $y=0.01$ as follows:
\begin{eqnarray}
\frac{\Delta I_{2}}{\Delta I_{1}} & \approx & -y \, \theta_{e} \, \approx  -0.03 \% \,  .
\end{eqnarray}
In the Rayleigh--Jeans region the multiple scattering contribution is safely neglected.

\section{DISCUSSION AND CONCLUDING REMARKS}

  From the results presented in the previous section it is clear that the multiple scattering contribution $\Delta I_{2}$ is very small compared with the single scattering contribution $\Delta I_{1}$.  For high-temperature galaxy clusters of $k_{B} T_{e} \approx 15$keV, we obtain the ratio $\Delta I_{2}/\Delta I_{1} \approx -0.3\%$ at $X=5$.  In the Rayleigh--Jeans region we have $\Delta I_{2}/\Delta I_{1} \approx -0.03\%$.  Therefore it is concluded that the multiple scattering contribution to the thermal Sunyaev-Zel'dovich effect for galaxy clusters can be safely neglected.  The reader is therefore referred to the previous four papers in this series of papers which deal with the single scattering contribution in detail.

\acknowledgements

This work is financially supported in part by the Grant-in-Aid of Japanese Ministry of Education, Science, Sports, and Culture under the contract \#10640289.

\newpage


\references{} 
\reference{} Arnaud, K. A., Mushotzky, R. F., Ezawa, H., Fukazawa, Y., Ohashi, T., Bautz, M. W., Crewe, G. B., Gendreau, K. C., Yamashita, K., Kamata, Y., \& Akimoto, F. 1994, ApJ, 436, L67
\reference{} Audit, E., \& Simmons, J. F. L. 1999, MNRAS, 305, L27
\reference{} Berestetskii, V. B., Lifshitz, E. M., \& Pitaevskii, L. P. 1982, $Quantum$ $Electrodynamics$ (Oxford: Pergamon)
\reference{} Birkinshaw, M. 1979, MNRAS, 187, 847
\reference{} Birkinshaw, M. 1999, Physics Reports, 310, 97
\reference{} Birkinshaw, M., \& Hughes, J., P. 1994, ApJ, 420, 33
\reference{} Birkinshaw, M., Hughes, J. P., \& Arnaud, K. A. 1991, ApJ, 379, 466
\reference{} Buchler, J. R., \& Yueh, W. R. 1976, ApJ, 210, 440
\reference{} Cavaliere, A., Danese, L., \& De Zotti, G. 1979, A\&A, 75, 322
\reference{} Challinor, A., Ford, M., \& Lasenby, A., 1999, MNRAS in press
\reference{} Challinor, A., \& Lasenby, A., 1998, ApJ, 499, 1
\reference{} Challinor, A., \& Lasenby, A., 1999, ApJ, 510, 930
\reference{} David, L. P., Slyz, A., Jones, C., Forman, W., \& Vrtilek, S. D. 1993, ApJ, 412, 479
\reference{} Furuzawa, A., Tawara, Y., Kunieda, H., Yamashita, K., Sonobe, T., Tanaka, Y., \& Mushotzky, R. 1998, ApJ, 504, 35
\reference{} Gunn, J. E. 1978, in Observational Cosmology, 1, ed. A. Maeder, L. Martinet \& G. Tammann (Sauverny: Geneva Obs.)
\reference{} Hansen, E., \& Lilje, P. B. 1999, MNRAS, 306, 153
\reference{} Herbig, T., Lawrence, C. R., Readhead, A. C. S., \& Gulkus, S. 1995, ApJ, 449, L5
\reference{} Holzapfel, W. L. et al. 1997, ApJ, 480, 449
\reference{} Itoh, N., Kohyama, Y., \& Nozawa, S. 1998, ApJ, 502, 7
\reference{} Itoh, N., Nozawa, S., \& Kohyama, Y. 2000, ApJ in press
\reference{} Itoh, N., Sakamoto, T., Kusano, S., Nozawa, S., \& Kohyama, Y. 2000, ApJS in press
\reference{} Jones, M. 1995, Astrophys. Lett. Commun., 6, 347
\reference{} Komatsu, E., Kitayama, T., Suto, Y., Hattori, M., Kawabe, R., Matsuo, H., Schindler, S., \& Yoshikawa, K. 1999, ApJ, 516, L1
\reference{} Kompaneets, A. S. 1957, Soviet Physics JETP, 4, 730
\reference{} Markevitch, M. 1998, ApJ, 504, 27
\reference{} Markevitch, M., Mushotzky, R., Inoue, H., Yamashita, K., Furuzawa, A., \& Tawara, Y. 1996, ApJ, 456, 437
\reference{} Markevitch, M., Yamashita, K., Furuzawa, A., \& Tawara, Y. 1994, ApJ, 436, L71
\reference{} Myers, S. T., Baker, J. E., Readhead, A. C. S., \& Herbig, T. 1995, preprint
\reference{} Mushotzky, R. F., \& Scharf, C. A. 1997, ApJ, 482, L13
\reference{} Nozawa, S., Itoh, N., Kawana, Y. \& Kohyama, Y. 2000, ApJ in press
\reference{} Nozawa, S., Itoh, N., \& Kohyama, Y. 1998a, ApJ, 507, 530
\reference{} Nozawa, S., Itoh, N., \& Kohyama, Y. 1998b, ApJ, 508, 17
\reference{} Rephaeli, Y. 1995, ApJ, 445, 33
\reference{} Rephaeli. Y., \& Yankovitch, D. 1997, ApJ, 481, L55
\reference{} Sazonov, S. Y., \& Sunyaev, R. A. 1998a, ApJ, 508, 1
\reference{} Sazonov, S. Y., \& Sunyaev, R. A. 1998b, Astronomy Letters 24, 553
\reference{} Sazonov, S. Y., \& Sunyaev, R. A. 1999, MNRAS in press
\reference{} Silk, J. I., \& White, S. D. M. 1978, ApJ, 226, L103
\reference{} Stebbins, A., 1997, preprint astro-ph/9705178
\reference{} Sunyaev, R. A., \& Zel'dovich, Ya. B. 1972, Comments Astrophys. Space Sci., 4, 173
\reference{} Sunyaev, R. A., \& Zel'dovich, Ya. B. 1980a, ARA\&A, 18, 537
\reference{} Sunyaev, R. A., \& Zel'dovich, Ya. B. 1980b, MNRAS, 190, 413
\reference{} Sunyaev, R. A., \& Zel'dovich, Ya. B. 1981, Astrophysics and Space Physics Reviews, 1, 1
\reference{} Tucker, W., Blanco, P., Rappoport, S., David, L., Fabricant, D., Falco, E. E., Forman, W., Dressler, A., \& Ramella, M. 1998, ApJ, 496, L5
\reference{} Weymann, R. 1965, Phys. Fluid, 8, 2112
\reference{} Zel'dovich, Ya. B., \& Sunyaev, R. A. 1969, Astrophys. Space Sci., 4, 301


\newpage
\centerline{\bf \large Figure Captions}

\begin{itemize}

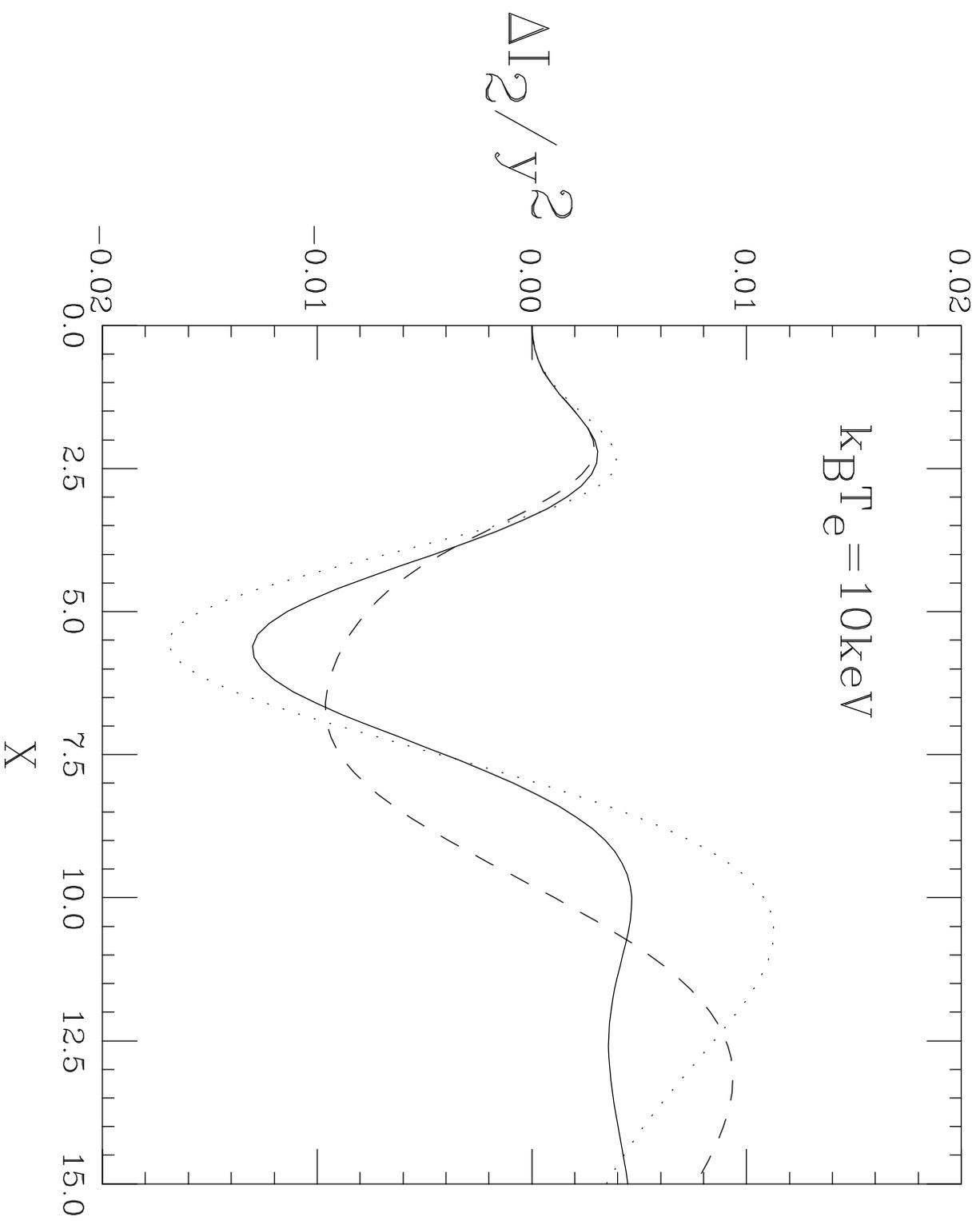
\item Figure 1. The spectral intensity distortion $\Delta I_{2}/y^{2}$ as a function of $X$ for the case $k_{B} T_{e}$ = 10keV.  The dotted curve shows the contribution up to $Z_{0}$.  The dashed curve shows the contribution up to $\theta_{e} Z_{1}$.  The solid curve shows the full contribution up to $\theta_{e}^{2} Z_{2}$.

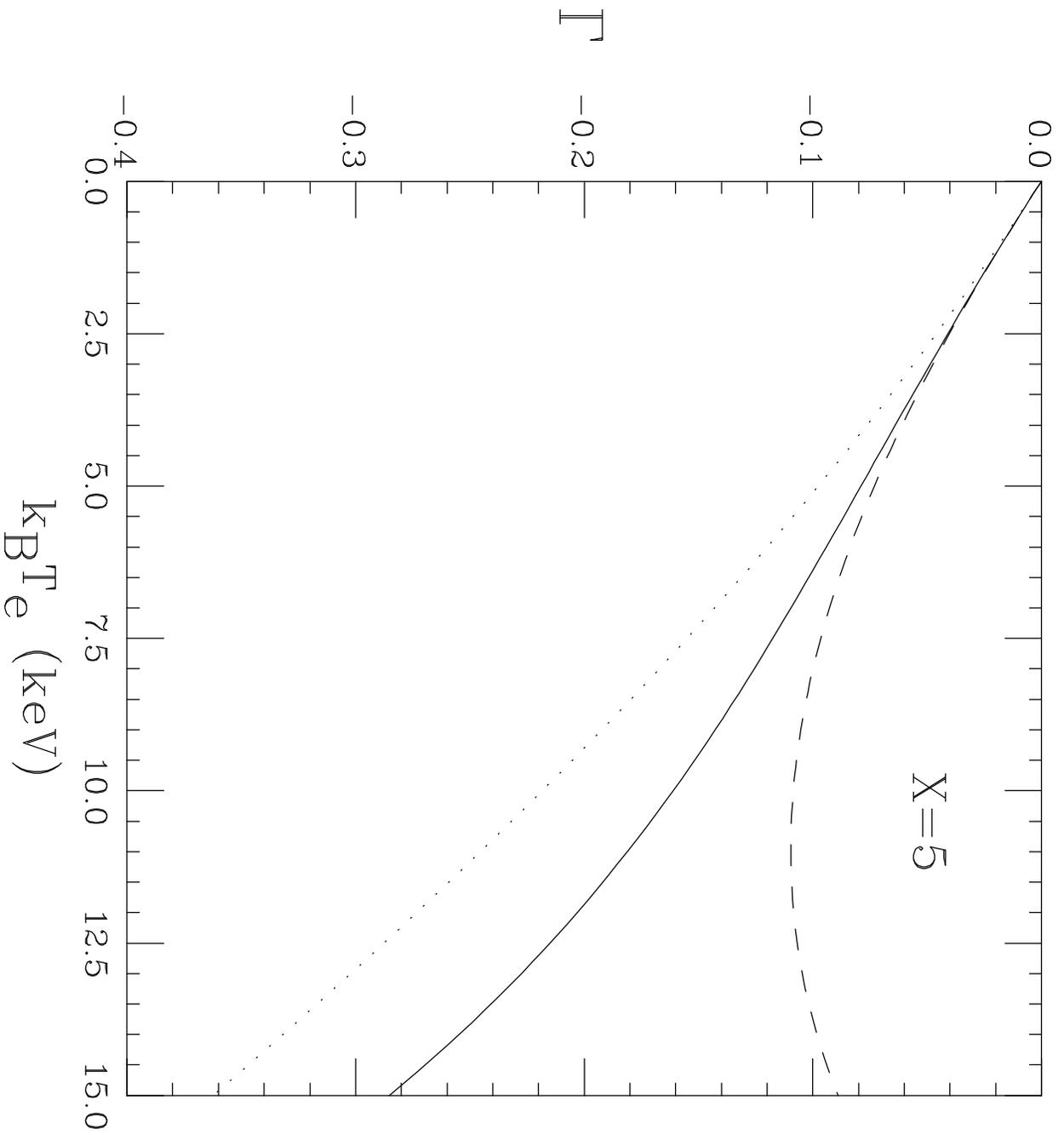
\item Figure 2. The ratio $\Gamma$ as a function of $k_{B} T_{e}$ for a fixed value of $X=5$.  The dotted curve shows the contribution up to $Z_{0}$.  The dashed curve shows the contribution up to $\theta_{e} Z_{1}$.  The solid curve shows the full contribution up to $\theta_{e}^{2} Z_{2}$.

\end{itemize}

\end{document}